\documentstyle[aps,prl,preprint,floats,epsfig]{revtex}

\def\B{\relax\ifmmode{B}\else{$B$}\fi}
\def\DS{\relax\ifmmode{D^{*}}\else{$D^{*}$}\fi}
\def\D {\relax\ifmmode{D}\else{$D$}\fi}
\def\PI{\relax\ifmmode{\pi}\else{$\pi$}\fi}
\def\USSSS{\relax\ifmmode{\Upsilon{ (4S)}}
	\else{$\Upsilon{ (4S)}$}\fi}
\def\DSP{\relax\ifmmode{D^{*+}}\else{$D^{*+}$}\fi}
\def\DSPM{\relax\ifmmode{D^{*\pm}}\else{$D^{*\pm}$}\fi}
\def\DSZ {\relax\ifmmode{D^{*0}}\else{$D^{*0}$}\fi}
\def\PIZ{\relax\ifmmode{\pi^0}\else{$\pi^0$}\fi}
\def\PIM{\relax\ifmmode{\pi^-}\else{$\pi^-$}\fi}
\def\PPP{\relax\ifmmode{\pi_f^-\pi_s^+}\else{$\pi_f^-\pi_s^+$}\fi}
\def\PZP{\relax\ifmmode{\pi_f^0\pi_s^+}\else{$\pi_f^0\pi_s^+$}\fi}
\def\PPZ{\relax\ifmmode{\pi_f^-\pi_s^0}\else{$\pi_f^-\pi_s^0$}\fi}
\newcommand{\ensuremath}[1]{\relax\ifmmode{#1}\else{$#1$}\fi}
\newcommand{\DSDP}{\ensuremath{\DS\!\to\!\D\PI}}
\newcommand{\BDSPPM}{\ensuremath{\overline{B}^0\!\to\!\DSP\,\PIM}}
\newcommand{\BDSZPM}{\ensuremath{\B^-\!\to\!\DSZ\,\PIM}}
\newcommand{\BDSP}{\ensuremath{\B\!\to\!\DS\,\PI}}
\newcommand{\BDSK}{\ensuremath{\B\!\to\!\DS\,K}}
\newcommand{\BBbar}{\ensuremath{B\overline{B}}}
\newcommand{\NBBbar}{\ensuremath{(3.27 \pm 0.06) \times 10^6}}
\newcommand{\valBZ}{\ensuremath{(2.81 \pm 0.11 \pm 0.21 \pm 0.05) \times 10^{-3}}}
\newcommand{\valBM}{\ensuremath{(4.34 \pm 0.33 \pm 0.34 \pm 0.18) \times 10^{-3}}}
\newcommand{\rmBB}{\ensuremath{{B}\overline{B}}}
\newcommand{\PISL}{\ensuremath{\PI_s}}

\newcommand{\PIFS}{\ensuremath{\PI_f}}

\newcommand{\CB}{\ensuremath{\cos{\theta^*_f}}}
\newcommand{\CDS}{\ensuremath{\cos{\theta^*_ s}}}

\begin{document}

\preprint{\tighten\vbox{\hbox{CLNS 97/1485}
	                \hbox{CLEO 97-11}
}}

\title{A New Measurement of $B \to D^* \pi$ Branching Fractions}  

\author{CLEO Collaboration}
\date{\today}

\maketitle
\tighten

\begin{abstract} 
The decays $\Upsilon(4S)\!\to\!B\overline{B}$, followed by
$B\!\to\!D^{*}\pi$ and $D^{*}\!\to\!D\pi$, 
permit reconstruction of all kinematic
quantities that describe the sequence
without reconstruction of the $D$, with reasonably
low backgrounds.
Using an integrated $e^+e^-$ luminosity of 
3.1 ${\rm fb}^{-1}$ accumulated at the $\Upsilon$(4S)
by the CLEO-II detector,
we report measurements of 
${\cal B}(\BDSPPM) = \valBZ$ and
${\cal B}(\BDSZPM) = \valBM$.

\end{abstract}

{
\renewcommand{\thefootnote}{\fnsymbol{footnote}}

\newpage
\begin{center}
G.~Brandenburg,$^{1}$ R.~A.~Briere,$^{1}$ Y.~S.~Gao,$^{1}$
D.~Y.-J.~Kim,$^{1}$ R.~Wilson,$^{1}$ H.~Yamamoto,$^{1}$
T.~E.~Browder,$^{2}$ F.~Li,$^{2}$ Y.~Li,$^{2}$
J.~L.~Rodriguez,$^{2}$
% ...
\end{center}
 
\small
\begin{center}
$^{1}${Harvard University, Cambridge, Massachusetts 02138}\\
$^{2}${University of Hawaii at Manoa, Honolulu, Hawaii 96822}\\
% ...
\end{center}

\newpage
\setcounter{footnote}{0}
}


The study of \B\ decays to exclusively hadronic final states has been
limited because samples in available data are small.  In this paper we
employ a technique, a ``partial reconstruction,'' that can increase
the acceptance of the sequence $\Upsilon$(4S)$\to\!B\overline{B}$,
$B\!\to\!D^{*}\pi$, $D^{*}\!\to\!D\pi$, by one
order of magnitude with respect to the more usual technique, 
``full reconstruction,'' where all particles in the final state 
are reconstructed.  For example, in a recent analysis~\cite{bigb97} 
using the latter technique, $248$ out of 
$\sim\!8700$ possible \BDSPPM\ decays
were reconstructed; in this letter, we report the reconstruction
of $\sim\!2600$ \BDSPPM\ from the same set of data.  We report on
the measurement of two of the \BDSP\ branching 
fractions with partial reconstruction, and we
probe the factorization hypothesis.   
The partial reconstruction might
enable an interesting sensitivity to a small $CP$ asymmetry in
\BDSPPM\ decays~\cite{cp}.

Both the CLEO~\cite{bigb97,CLEO:BigB} and 
ARGUS~\cite{Argus:Full} collaborations reported 
measurements of \BDSP\ based on the full reconstruction technique.
In the analysis of data presented in this letter,
all kinematic quantities that describe the
decay chain $B \rightarrow D^* \pi_f$, $D^* \rightarrow D \pi_s$ 
are reconstructed
from measurements of the three-momenta of the two
pions, one fast ($\pi_f$) and one slow ($\pi_s$), $\vec{p}_f$ and $\vec{p}_s$;
the $D$ from \DS\ decay is undetected, which yields an order of magnitude
increase in acceptance over full reconstruction, and removes systematic
uncertainty introduced by $D$ branching fractions.
 
The basic idea was described in~\cite{CLEO:sheldon}: a $B$ from
$\Upsilon(4S)$ decay is nearly at rest and the energy release in 
the $D^* \rightarrow D \pi$ decay is small, so the decay products
, $\pi_s$ and $\pi_f$, are nearly back to back.
The smearing introduced in~\cite{CLEO:sheldon} by neglect of the detailed
kinematics of the decay sequence is much larger than
the smearing caused by errors in either the measurement
of the pion momenta, or by the error in knowledge of the
magnitude of the initial $B$ momentum.  Complete evaluation
of the detailed kinematics leads to a significant improvement
in the description of the shape of the signal, the shape
of the background, and rejection of the background.

To fully describe the
kinematics of the decay, twenty parameters are required:
four for each four-vector of the five particles: $B$, $D^*$, $\pi_f$,
$D$, and $\pi_s$. Energy-momentum conservation can be
applied twice, in the $B \rightarrow D^* \pi_f$ and 
$D^* \rightarrow D \pi_s$ decays,
yielding eight equations; the masses of the five particles
can be assumed; and the center-of-mass energy of the
$e^+e^-$ collisions can be used to obtain the magnitude
of the three-momentum of the initial $B$.
The six free parameters that remain describe the kinematics
of the decay sequence.  These can be thought of as
six angles: two that describe the \B\ direction, two angles
($\theta^*_f$,$\phi^*_f$)
that describe the direction of the $\pi_f$ in the $B$
rest frame, and two angles ($\theta^*_s$,$\phi^*_s$)
that describe the direction
of the $\pi_s$ in the $D^*$ rest frame.  We evaluate
those six angles from the measurement of the three components
of the $\pi_f$ momentum and the three components of the $\pi_s$
momentum.  

The angles that
provide effective discrimination between signal and
background are $\theta^*_f$ and $\theta^*_s$, for which
the explicit expressions are:
\begin{eqnarray}
\cos\theta^*_f & = & \frac{-\beta_B(E^*_f - E^*_{D^*})}{2 P^*_f} + 
\frac{{\left| \vec{p}_f \right| }^2 - {\left| P_{D^*} \right|}^2}
{2\gamma_B^2 \beta_B M_B P^*_f} {\rm\ \ \;\;and} \\
\cos\theta^*_s & = & \frac{-\beta_{D^*}(E^*_s - E^*_{D})}{2 P^*_s} + 
\frac{{\left| \vec{p}_s \right|}^2 - {\left| P_{D} \right|}^2}
{2\gamma_{D^*}^2 \beta_{D^*} M_{D^*} P^*_s},
\end{eqnarray}
where $E^*_f$, $E^*_{D^*}$ and $P^*_f$ are the energy and momentum of the
$\pi_f$ and $D^*$ in the $B$ center of mass;
$E^*_s$, $E^*_{D}$ and $P^*_s$ are the energy and momentum of the
$\pi_s$ and $D$ in the $D^*$ center of mass;
$\gamma_{B(D^*)}$, $\beta_{B(D^*)}$ and $M_{B(D^*)}$ are the Lorentz factor,
the velocity and the mass of the $B(D^*)$ in the lab frame.
The magnitude of the $D^*$ and $D$ momenta in the lab frame, 
$\left| P_{D^*} \right|$ and $\left| P_{D} \right|$, are determined
by applying energy-momentum conservation in the decay chain.
For signal, the magnitudes of these cosines will tend
to fall into the `physical' region; less than one.
The signal distribution will be uniform
in $\cos\theta^*_f$ (because the $B$ has spin 0),
and as $\cos^2\theta^*_s$ (because the $D^*$ has helicity 0),
before consideration of detector acceptance, efficiency,
and resolution.  Detector resolution sometimes pushes
signal events into the `non-physical' region, where the magnitude
of one or both of the cosines exceeds unity.
Backgrounds usually
fall into the non-physical region. The variables
$\cos\theta^*_f$ and $\cos\theta^*_s$ tend to depend
linearly on $|\vec{p}_f|$ and $|\vec{p}_s|$ once the dependence
of $\left| P_{D^*} \right|$ and $\left| P_{D} \right|$ on these
variables is included.

The angle between the plane of the $B \rightarrow D^* \pi_f$ decay 
and the plane of the $D^* \rightarrow D \pi_s$ decay, 
$\phi = \phi^*_f - \phi^*_s$, is reconstructed
in the following manner.  In the lab frame, the $D^*$ direction
must lie on a small cone of angle $\alpha_f$ around the direction
\emph{opposite} to the $\pi_f$.  Simultaneously, the $D^*$ must
\emph{also} lie on a second small cone of angle $\alpha_s$
around the direction of the $\pi_s$. The expressions for these
angles are:
\begin{equation}
\cos\alpha_f = \frac{M_B^2 - M_{D^*}^2 - M_\pi^2}
{2\left| P_{D^*} \right|\left| \vec{p}_f \right|} - 
\frac{1}{\beta_{D^*}\beta_f} {\rm\ \ \;\;and\;\;}
\cos\alpha_s = -\frac{M_{D^*}^2 + M_\pi^2 - M_D^2}
{2\left| P_{D^*} \right|\left| \vec{p}_s \right|} + 
\frac{1}{\beta_{D^*}\beta_s},
\end{equation}
where the momenta and velocities are measured in the lab frame.
The decay kinematics limit $\alpha_f \leq 0.14$ and
$\alpha_s \leq 0.28$.
Intersection of
these two cones determines the $D^*$ directions, of
which in practice there are two: a so-called quadratic
ambiguity.  For both $D^*$ directions:
\begin{equation}
\cos\phi= \frac{\cos\delta - \cos\alpha_f \cos\alpha_s}
{\sin\alpha_f \sin\alpha_s},
\end{equation}
where $\delta$ is the angle between $\vec{p}_s$ and the direction opposite
to $\vec{p}_f$.
For most signal events $|\cos\phi|<1$, or `physical'.  
Signal events with imperfect measurement of the pion momenta, 
as well as non-signal events, can result in $|\cos\phi|>1$, 
in most cases because $\delta>\alpha_f+\alpha_s$.
 
The data used in this analysis were selected from 
hadronic events produced in $e^+e^-$
annihilations at the Cornell Electron Storage Ring (CESR).  
The data sample consists of 3.1 ${\rm fb}^{-1}$ 
collected with the CLEO-II detector~\cite{Detector} at the 
\USSSS\ resonance (referred to as 
`on-resonance') and 1.6 ${\rm fb}^{-1}$ at a 
center-of-mass energy just below the threshold for
production of \rmBB\ pairs (referred to as 
`off-resonance').  The on-resonance data 
correspond to \NBBbar\ \rmBB\ pairs.  
The off-resonance data are used 
to model the background from non-\rmBB\ decays.  

Charged pions that are consistent with production
at the $e^+e^-$ annihilation position and that penetrate
all layers of the CLEO II tracking system are identified
by means of time-of-flight, specific ionization, and shower
development in the CsI calorimeter and surrounding muon
identifier.  Neutral pions are reconstructed primarily from
information in the CsI calorimeter~\cite{Detector}.

Events with two pions are classified according to the net charge,
which is $0$ or $\pm1$ for signal.
The fast pion is charged, but the slow pion can be either
charged (\PPP) or neutral (\PPZ).
Only \DSPM\ decays yield slow charged pions, but
slow neutral pions are produced from both
\DSZ\ and \DSPM\ decays, and so the
\PPZ\ sample will contain contributions from both
\BDSPPM\ and \BDSZPM.
We further require that events satisfy
the ``$D^*$ cone overlap requirement'': 
$\left| \cos\delta - \cos\alpha_f \cos\alpha_s \right| <
\sin\alpha_f \sin\alpha_s + 0.02$, which allows for detector resolution.

Some events satisfy all requirements two or more times,
usually through combinations of one fast pion with several distinct
slow pions.  In signal Monte Carlo studies,
5\% (24\%) of \PPP\ (\PPZ) events have more than one possible 
slow charged (neutral) pion.
In \PPZ\ events, we select the neutral pion whose mass is 
closest to the nominal \PIZ\ mass and in \PPP\ events the 
two pion candidate with the smallest value of 
$\left| \cos\delta - \cos\alpha_f \cos\alpha_s \right|$ is selected.

The dominant sources of background are non-\rmBB\ events. 
The distribution of decay products in these events tends to be jet-like,
while in \rmBB\ events the decay products tend to be distributed uniformly
in angle.
To suppress non-\rmBB\ events, each candidate event must satisfy $R_2 < 0.275$,
where $R_{2}$ is the
ratio of the second Fox-Wolfram moment to the zeroth moment~\cite{FoxWolf}.
We also reject events where the momentum of any charged track exceeds
the maximum possible from a $B$ decay, 2.45 GeV/c.

To extract the branching fractions we perform a two-dimensional
fit in \CB\ and \CDS, where 
the fit region is $|\CB| < 1.65$ and $-1.6 < \CDS < 5.0$.
The \PPP\ and \PPZ\ data samples are fit 
simultaneously using the MINUIT~\cite{minuit} program.
The fitting function combines contributions from the \BDSP\ signal, 
other \B\ decays, and a fixed amount of non-\rmBB\
background as described below.

The non-\rmBB\ background shape and rate is determined from a
sample of off-resonance data, that has been scaled
for the relative luminosities and cross-sections between the on-resonance and 
off-resonance data samples.
The \CB\ and \CDS\ distributions in non-\rmBB\ events are primarily 
determined by the \PIFS\ and \PISL\
momentum spectra in those events.  Additionally, the
shape is affected by the $D^*$ cone overlap
requirement, which admits the most events when $\alpha_s$ is largest,
which occurs roughly when \CDS\ $\approx 0$.  
The shape of the background is thus roughly $\propto\sin^2\theta^*_s$,
which is the complement of the signal, $\propto\cos^2\theta^*_s$.

A large sample of simulated \rmBB\ events shows that
this background is dominated by modes that
are able to produce a fast pion candidate,
such as
$B \rightarrow D^{(\pm,0)}X$, where the $X$ system is 
predominantly $\pi$, $\rho$ or $\mu \nu_{\mu}$, and the $D^{(\pm,0)}$ 
can be in an excited state.  
The background distribution in $\cos\theta_f^*$ is determined by the
kinematics of the fast pion from the $B$ decay.
Slow pions are plentiful in these $\rmBB$ background samples.
When fast and slow pion candidates come from different $B$'s,
the resulting distribution in $\cos\theta_s^*$
resembles the non-\rmBB\ distribution.
When both candidates come from the same \B\ decay, the distribution
in $\theta_s^*$ and $\theta_f^*$ is distinctive, but unlike that of 
the signal: the branching ratios of modes that enter the final sample
in this manner are allowed to float in the final fit,
either constrained by a Gaussian to the central value and error
in~\cite{PDG96c}, or left
unconstrained, if no measurement is available.  The branching
fractions used in the \PPP\ and the \PPZ\ samples are constrained
to be equal.

One \B\ decay background mode is handled differently.  The 
Cabibbo-suppressed mode \BDSK\ is essentially
indistinguishable from \BDSP\ in the partial reconstruction. 
We assume that the ratio of branching fractions, 
${\cal B}(\BDSK)/{\cal B}(\BDSP)$, is given by the ratio of the
decay constants for kaons and pions, $f_K/f_\pi$, the ratio of the 
CKM matrix elements, $V_{us}/V_{ud}$, and the ratio of form factors.
The product of these ratios is determined to 
be ($7.69 \pm 0.08$)\%~\cite{PDG96a,Neubert}.
The assumed \BDSK\ rate is subtracted, with adjustment
for acceptance.

\begin{figure}[p]
\psfig{figure=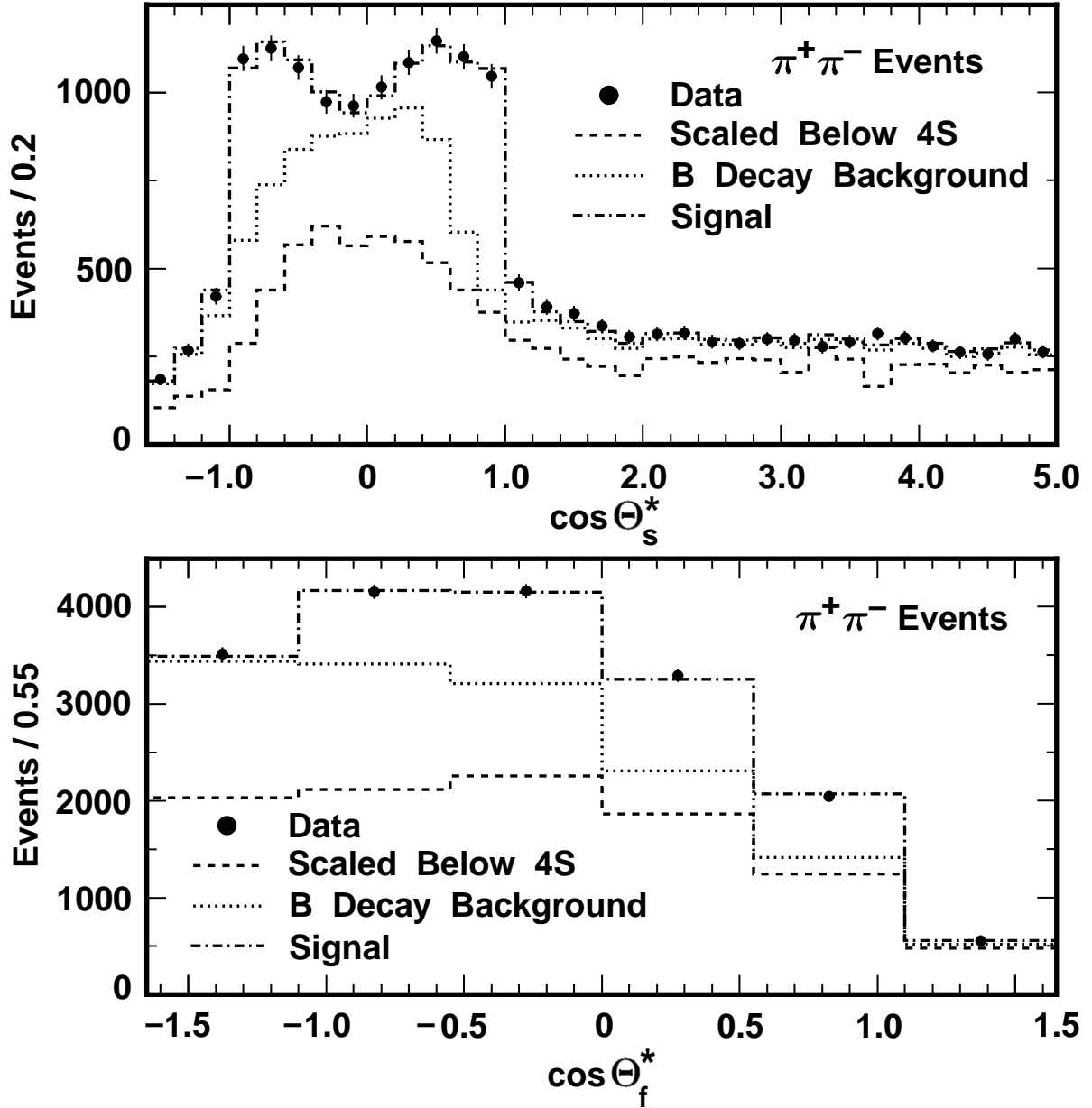,width=\textwidth}
\caption
{The projections of the data histogram in \CB\ and \CDS\ 
with the fitting function for the \PPP\ fit.}
\label{fig:wfig1}
\end{figure}

\begin{figure}[p]
\psfig{figure=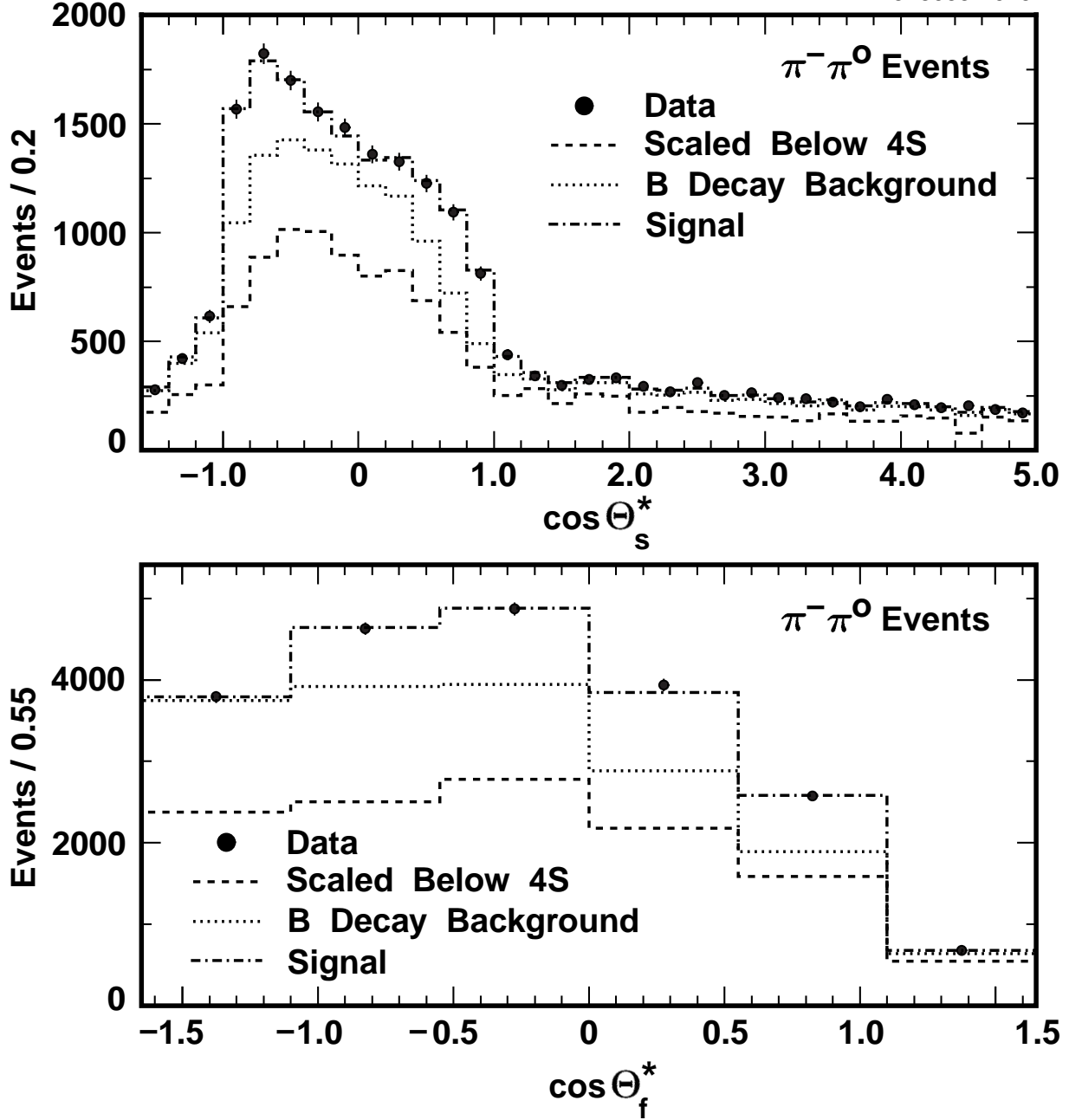,width=\textwidth}
\caption
{The projections of the data histogram in \CB\ and \CDS\ 
with the fitting function for the \PPZ\ fit.}
\label{fig:wfig2}
\end{figure}

\begin{figure}[p]
\psfig{figure=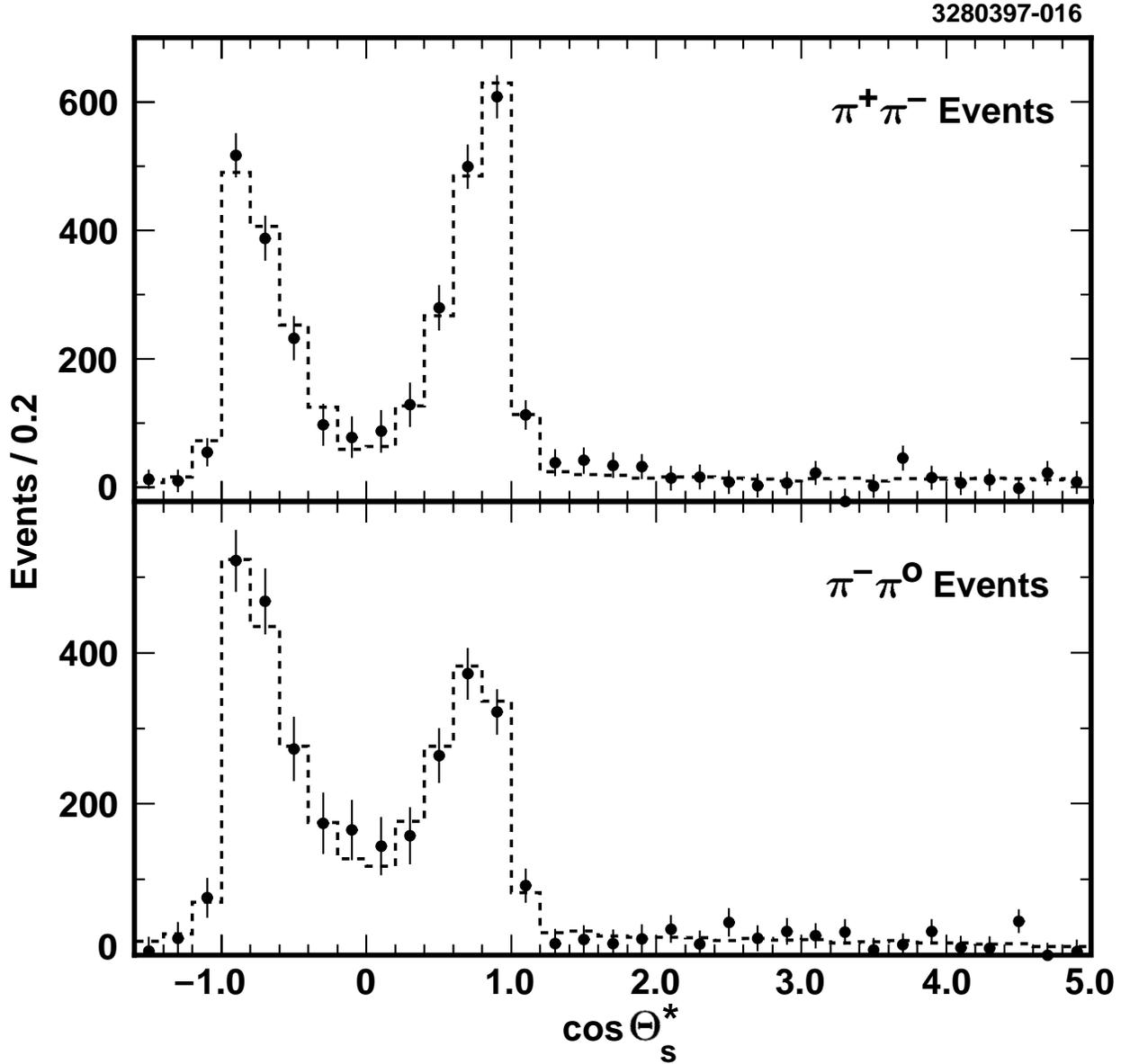,width=\textwidth}
\caption
{The background-subtracted projections of the data histogram in \CDS\ 
for the \PPP\ and \PPZ\ fits.  The dashed line is the signal shape.}
\label{fig:wfig3}
\end{figure}

The projections of the data and the fitting function in \CB\ and \CDS\
are shown in Fig.~\ref{fig:wfig1} for the \PPP\ fit and in 
Fig.~\ref{fig:wfig2} for the \PPZ\ fit.  The sidebands outside
the signal region tend to determine the background normalization,
and are fitted well by the background functions.
The sharp turn-on of signal at $\pm 1$ can be seen while the 
background distribution in \CDS\ shows the expected peaking
in the signal region due to the $D^*$ cone overlap requirement.
The confidence level for the \PPP(\PPZ) fit alone is 29\%(2\%).
No structure is observed in the residuals of the fit and
confidence level for the combined fit is 3\%.
The fitted number of signal events is given in Tab.~\ref{tab:yield} along
with the product of acceptance and efficiency and the 
relevant $D^*$ branching fraction.  
The background subtracted plots for the \PPP\ and \PPZ\ fits
for the \CDS\ projection are shown in Fig.~\ref{fig:wfig3}.  The
peaks are asymmetric because the acceptance functions 
for charged and neutral slow pions have momentum dependences.

The systematic uncertainty was determined to be 7.5\% for \BDSPPM\ and
8.3\% for \BDSZPM.  The error is dominated by uncertainties in
the slow pion reconstruction efficiency, \B\ decay background shape
and simulation of the $R_2$ requirement.  Additional errors come from the
uncertainty in the number of \rmBB\ pairs produced, signal shape smearing,
Monte Carlo statistics and the simulation of $\cos\delta$.

To convert from fitted yields to branching fractions we
use the value of \NBBbar\ \rmBB\ 
pairs produced and assume that the ratio of 
$B^+B^-$ to $B^0\overline{B}^0$ production ($f_{+-}/f_{00}$) is one.  
This is in agreement with the current CLEO measurement of
$f_{+-}\tau_{B^\pm}/f_{00}\tau_{B^0} = 1.15 \pm 0.17 \pm 0.06$ 
\cite{CLEO:wanke} and 
the value~\cite{PDG96c} for the ratio of lifetimes 
$\tau_{B^\pm}/\tau_{B^0} = 1.03 \pm 0.06$.  We find:

\begin{equation}
{\cal B}(\overline{B}^0 \to\ D^{*+} \pi^-) = \valBZ
\end{equation}
\begin{equation}
{\cal B}(B^- \to\ D^{*0} \pi^-) = \valBM
\end{equation}
\noindent
where the first error is statistical, the second is systematic, and
the third comes from the uncertainty in the \DSDP\ branching fractions.

To compare with the factorization hypothesis~\cite{BSW:1},
we take the ratio of charged to neutral branching fractions, 
in which the systematic uncertainties
due to the number of \BBbar\ events, the $R_2$ requirement, and the fast pion
reconstruction cancel.  The ratio is measured to be $r=1.55 \pm 0.14 \pm 0.15$.

An implementation of the factorization hypothesis\cite{BSW} predicts that 
$r$ is equal to $(1 + 1.29\ a_2/a_1)^2$.
The coefficient $a_1\!\approx\!1$ describes the `external spectator
amplitude,' where the $W$ hadronizes to a single pion,
and $a_2$ describes the internal, color-suppressed amplitudes, and
is expected to be rather smaller than 1.  The measurement of $r$
yields $a_2/a_1$ of $0.19 \pm 0.04 \pm 0.05$.
Another ratio, ${\cal B}(\overline{B}^0\!\to\!D^{*0} \pi^0)/
{\cal B}(\overline{B}^0\!\to\!D^{*+} \pi^-)$ is given by 
$0.84 \times (a_2/a_1)^2$
using the same model. From the results quoted above, the factorization
hypothesis predicts, in the absence of final state interactions,
 ${\cal B}(\overline{B}^0\!\to\!D^{*0} \pi^0) = 8.5 \times 10^{-5}$,
about five times smaller than the current~\cite{CLEO:yongsheng}
experimental limit.

We searched for the suppressed modes which produce a fast neutral
pion.  In \PZP\ events no signal was observed. The confidence level of the
fit was 73\% indicating good agreement between the background shape
and the data.
We limit the doubly CKM-suppressed
mode to ${\cal B}(B^-\!\to\!D^{*-} \pi^0) < 1.7 \times 10^{-4}$ at 90\%
confidence level.  For internal color-suppressed modes the superior background
rejection of the full reconstruction technique~\cite{CLEO:yongsheng} 
leads to better sensitivity, except in the
case of $\overline{B}^0\!\to\!D^{*0} \eta^\prime$.
We set a limit of 
${\cal B}(\overline{B}^0\!\to\!D^{*0} \eta^\prime) < 14 \times 10^{-4}$ at 90\%
confidence level.  The confidence level of the fit was 10\%. 

We gratefully acknowledge the effort of the CESR staff in providing us with
excellent luminosity and running conditions.
This work was supported by 
the National Science Foundation,
the U.S. Department of Energy,
the Heisenberg Foundation,  
the Alexander von Humboldt Stiftung,
%
Research Corporation,
the Natural Sciences and Engineering Research Council of Canada,
and the A.P. Sloan Foundation.

\begin{table}[p]
\caption{The yield of signal events from the fits.  The D* branching fractions
are not included in the calculation of acceptance times efficiency.}
\vspace{.5em}
\begin{center}
\begin{tabular}{lccc}
Mode & Yield & Acc. $\times$ Eff. & ${\cal B}$($D^*\!\to\!D\pi$) \\
\hline \hline
$\overline{B}^0\!\to\!D^{*+} \pi^-  ; D^{*+}\!\to\!D^0 \pi^+ $ & $2612\pm102$ & 0.42 & 68.3\% \\
$\overline{B}^0\!\to\!D^{*+} \pi^-  ; D^{*+}\!\to\!D^+ \pi^0 $ & $513\pm21$ & 0.18 & 30.6\% \\
$B^-\!\to\!D^{*0} \pi^-  ; D^{*0}\!\to\!D^0 \pi^0 $ & $1560\pm115$ & 0.18 & 61.9\% \\
\end{tabular}
\end{center}
\label{tab:yield}
\end{table}

\end{document}